\documentstyle[pre,aps,epsf,multicol]{revtex}

\newcommand{\av}[2]{\left\langle#1\right\rangle_{#2}}

\newcommand{\eg}{{\rm e.g.,}}

\newcommand{\epsg}{\epsilon_{\rm g}}

\newcommand{\half}{{1 \over 2}}
\newcommand{\ie}{{\rm i.e.,}}

\newcommand{\n}{p}

\newcommand{\qeff}{q_{\rm eff}}

\newcommand{\reff}{R_{\rm eff}}

\newcommand{\sgn}{{\rm sgn}}

\newcommand{\T}{^{{\rm T}}}
\newcommand{\thetnplus}{\th{\n+1}}

\newcommand{\thn}{\th{\n}}
\newcommand{\th}[1]{\Theta^{(#1)}}

\newcommand{\wvec}{{\bf w}}

\newcommand{\xvec}{{\bf \x}}
\newcommand{\x}{x}
\newcommand{\y}{y}

\def\(#1){(\ref{#1})}
\def\$$#1\$${\begin{equation} #1 \end{equation}}
\def\$[#1\$]{\begin{equation} #1 \end{equation}}

\newcommand{\ba}{\begin{eqnarray}}
\newcommand{\ea}{\end{eqnarray}}
\newcommand{\binomial}[2]{
\let\arraystretchsave\arraystretch
\def\arraystretch{0.7}
\left(\!\!\begin{array}{c}#1\\#2\end{array}\!\!\right)
\let\arraystretch\arraystretchsave
}

\begin{document}

\draft
 %\preprint{Preprint Version 1.0; \today}

\title{%A quantitative analysis of query
 %Query
 %learning for maximum
 %information gain in a multi-layer neural network}
Learning from minimum entropy queries in a large committee machine
}

\author{Peter Sollich}
\address{Department of Physics, University of Edinburgh,
         Kings Buildings, Mayfield Road, Edinburgh EH9 3JZ, U.K.}

%\twocolumn

\maketitle

\begin{abstract}
 In supervised learning, the redundancy contained in random examples can
be avoided by learning from queries.  Using statistical mechanics, we
study learning from minimum entropy queries in a large tree-committee
machine.  The generalization error decreases exponentially with the
number of training examples, providing a significant improvement over
the algebraic decay for random examples.  The connection between entropy
and generalization error in multi-layer networks is discussed, and a
computationally cheap algorithm for constructing queries is suggested
and analysed.
 \end{abstract}

\vspace*{4mm}
\centerline{\small To appear in {\bf Physical Review E}, Rapid Communications.
%}
%\pacs{
}

\begin{multicols}{2}

 %\section{Introduction}
\label{sec:tcm_intro}
 In supervised learning of input-output mappings, the traditional
approach has been to study generalization from random examples.
However, random examples contain redundant information, and
generalization performance can thus be improved by {\em query learning},
where each new training input is selected on the basis of the existing
training data to be most `useful' in some specified sense.  In this
paper, we consider {\em minimum entropy queries}, defined by maximizing
the most common
measure of `usefulness', namely, the expected entropy decrease (or
information gain).  In order to achieve optimal generalization
performance, the theoretically optimal choice of queries would of course
be based on a direct minimization of the generalization error, and not
on maximization of the entropy decrease.  However, the generalization
error is not in general accessible as an objective function for query
selection, while the expected entropy decrease of a query can often be
determined fairly easily.  Since decrease in entropy and generalization
error are normally correlated (see, \eg\ Refs.~\cite{Freundetal93,Sollich94}),
minimizing
entropy therefore provides a practical method for achieving near-optimal
generalization performance by query learning.

The generalization performance achieved by minimum entropy
queries is by now well understood for single-layer neural networks such
as linear and binary
perceptrons~\cite{Freundetal93,Sollich94,Seungetal92b}.  For multi-layer
networks, which are much more widely used in practical applications,
several heuristic algorithms for query learning have been
proposed  (see \eg\ Refs.~\cite{Baum91,Hwangetal91}).
 % proposed.
While such heuristic approaches can
demonstrate the power of query learning, they are hard to generalize to
situations other than the ones for which they have been designed, and
they cannot easily be compared with more traditional techniques for
query selection such as optimal experimental design.
Furthermore, the existing analyses of such
algorithms have been carried out within the framework of PAC (probably
approximately correct) learning, yielding worst case bounds which
do not necessarily represent average case behaviour.
In this paper we therefore analyse the average generalization
performance achieved by query learning in a multi-layer network, using
the tools of statistical mechanics.

 % In particular, we consider
 %query learning in a large {\em tree-committee machine} (TCM), with noise free
 %training data generated by a teacher network of the same architecture.
 %The details of the model are explained in the next section. In
 %section~\ref{sec:calcul_exact_msse_queries}, we then outline the calculation
 %of the main quantity of interest, the average generalization error
 %$\epsg$ as a function of the (normalized) number of training examples,
 %$\alpha$. The results are compared to existing analyses of learning from
 %random examples in a TCM and related to
 %corresponding results for the binary perceptron.
 %We also discuss the relationship between
 %information gain and generalization error in the TCM. In
 %section~\ref{sec:calcul_appr_msse_queries}, we analyse a computationally
 %cheap algorithm for constructing approximate minimum entropy
 %queries, and find that it achieves generalization performance even
 %slightly superior to that of exact minimum entropy queries.
 %In section~\ref{sec:tcm_conclusions}, we summarize and discuss our results
 %and offer our conclusions regarding the potential for practical
 %applications of query learning in multi-layer neural networks.

 %\section{The model}
 %\subsection{Tree-committee machine}

We focus on one of the simplest multi-layer neural networks,
namely, the tree-committee machine (TCM).
A TCM is a two-layer network with $N$
input units, $K$ hidden units and one output unit.  The `receptive
fields' of the individual hidden units do not overlap,
 %, and each hidden
 %units calculates the sign of a linear combination (with real
 %coefficients) of the $N/K$ input components to which it is connected.
 %The output unit then calculates the sign of the sum of all the hidden
 %unit outputs.  A TCM therefore effectively has
and all the weights from the
hidden to the output layer are fixed to one.  The output $\y$
for a given input vector $\xvec$ is therefore
 \$[
\y =
 %
 %f(\xvec;\wvec) =
 %
 \sgn \left( {1\over \sqrt{K} } \sum_{i=1}^K \sigma_i\right) \qquad
\sigma_i=\sgn\left(\sqrt{{K\over N}}\xvec_i\T\wvec_i\right)
\label{tcm_output}
 \$]
 where the $\sigma_i$ are the outputs of the hidden units, $\wvec_i$
their weight vectors and $\xvec\T=(\xvec_1\T,\ldots,\xvec_K\T)$ with
$\xvec_i$ containing the $N/K$ real-valued inputs which hidden unit $i$
receives.
 %~\cite{Kodd_footnote}.
 The $N$ components of the $K$ $(N/K)$-dimensional hidden unit weight
vectors $\wvec_i$, which we denote collectively by $\wvec$, form the
adjustable parameters of a TCM.  Without loss of generality,
the weight vectors are assumed to be normalized
to $\wvec_i^2=N/K$, corresponding roughly to individual weights of $O(1)$.
 % this ensures
 %that the arguments of the $\sgn$-functions in~\(tcm_output) are of order
 %unity for a typical input vector $\xvec$ of length $\xvec^2=N$.
 %
 % We shall restrict our analysis to the case where both the input space
 %dimension and the number of hidden units are large ($N\to\infty$,
 %$K\to\infty$), assuming that each hidden unit is connected to a large
 %number of inputs, \ie\ $N/K\gg 1$.

As our training algorithm we take
(zero temperature) Gibbs learning, which generates at random any TCM (in
the following referred to as a `student') which predicts all the
training outputs in a given set of $p$ training examples
$\thn=\{(\xvec^\mu,\y^\mu), \mu=1\ldots p\}$ correctly.  We take the
problem to be perfectly learnable, which means that the outputs $\y^\mu$
corresponding to the inputs $\xvec^\mu$ are generated by a `teacher' TCM
with the same architecture as the student but with different, unknown
weights $\wvec^0$.  It is further assumed that there is no noise on the
training examples.  For learning from random examples, the training
inputs $\xvec^\mu$ are sampled randomly from a distribution
$P_0(\xvec)$.  Since the output~\(tcm_output) of a TCM is independent of
the length of the hidden unit input vectors $\xvec_i$, we assume this
distribution $P_0(\xvec)$ to be uniform over all vectors
$\xvec\T=(\xvec_1\T,\ldots,\xvec_K\T)$ which obey the spherical
constraints $\xvec_i^2=N/K$.

For query learning, the training inputs
$\xvec^\mu$ are chosen
 to maximize the expected decrease of
 %information gain of the
 %student, as explained in the next section.
 %
 %\subsection{minimum entropy queries}
 %
 %We now explain how queries for minimum entropy in the TCM are
 %chosen. The information gain is defined as the decrease in
 the entropy
$S$ in the parameter space of the student. The entropy for a given
training set $\thn$ is defined as
\end{multicols}
\twocolumn
\noindent
 \$[
S(\thn)= - \int d\wvec P(\wvec|\thn) \ln P(\wvec|\thn).
\label{tcm_entropy_general}
 \$]
 For the Gibbs learning algorithm considered here, $P(\wvec|\thn)$
is uniform on the `version space', the space of all students satisfying
the spherical constraints $\wvec_i^2=N/K$ which
predict all training outputs correctly,
 %(and which satisfy the assumed
 %spherical constraints on the weight vectors, $\wvec_i^2=N/K$),
 and zero otherwise.  Denoting the version space volume by $V(\thn)$,
the entropy can thus simply be written as $S(\thn)=\ln V(\thn)$.  The
entropy decrease $\Delta S=S(\thn)-S(\thetnplus)$ resulting from the
addition of a new example $(\xvec^{p+1},\y^{p+1})$ to the existing
training set cannot be maximized directly, since it depends on the new
training output $\y^{p+1}$ generated by the unknown teacher.  Queries
are thus chosen to maximize the {\em expected} entropy decrease,
obtained by averaging over $\y^{p+1}$.  Assuming a uniform prior over
teachers, the probability of a certain teacher having produced the
training set $\thn$ is uniform over the version space and zero
otherwise.  The probability of obtaining output $\y^{p+1}=\pm 1$ given
input $\xvec^{p+1}$ is therefore simply
$v^\pm=V(\thetnplus)|_{\y^{p+1}=\pm 1}/V(\thn)$, the fraction of the
version space left over after the new example ($\xvec^{p+1},\y^{p+1}=\pm
1$) has been added~\cite{Seungetal92b}.  This gives the expected entropy
decrease
 \[
\av{\Delta S}{P(\y^{p+1}|\xvec^{p+1}, \thn)}=-v^+ \ln v^+ -v^- \ln v^-
 \]
 which attains its maximum value $\ln 2$ ($\equiv$ 1 bit) when
$v^\pm=\half$, \ie\ when the new input $\xvec^{p+1}$ {\em bisects} the
existing version space.  This is intuitively reasonable, since
$v^\pm=\half$ corresponds to maximum uncertainty about the new output
and hence to maximum information gain once this output is known.

Due to the complex geometry of the version space, the generation of
queries which achieve exact bisection is in general computationally
infeasible.  The `query by committee' algorithm~\cite{Seungetal92b}
provides a solution to this problem by first sampling
a `committee' of $2k$ students
 %~\cite{k_not_equal_K_footnote}
 from the Gibbs distribution $P(\wvec|\thn)$ and then using the fraction
of committee members which predict $+1$ or $-1$ for the output $\y$
corresponding to an input $\xvec$ as an approximation to the true
probability $P(\y=\pm 1|\xvec, \thn)=v^\pm$.  The condition
$v^\pm=\half$ is then approximated by the requirement that exactly $k$
of the committee members predict output $+1$ and the other $k$ predict
$-1$ for the new training input $\xvec^{p+1}$.  An
approximate minimum entropy query can thus be found by sampling
(or {\em filtering}) inputs from a stream of random inputs until this
condition is met.  The procedure is then repeated for each new query.
As $k\to\infty$, this algorithm approaches exact bisection,
and we focus on this limit in the following.

 %\section{Exact minimum entropy queries}
\label{sec:calcul_exact_msse_queries}

The main quantity of interest in our analysis is the generalization
error $\epsg$, defined as the probability that a given student TCM will
predict the output of the teacher incorrectly for a random test input
sampled from $P_0(\xvec)$.
We consider the thermodynamic limit $N\to\infty$ at constant number of
training examples per weight, $\alpha=p/N$, and focus on the case
of a large number of hidden units, $K\to\infty$ with $N/K\gg 1$.
The generalization error then takes the form~\cite{Schwarzeetal92}
 \$[
\epsg=(1/\pi)\arccos \reff
\label{tcm_generr}
 \$]
 where $\reff$ is an effective overlap parameter given by
 \[
 %\reff={1\over K} \sum_{i=1}^K {2\over\pi}\arcsin R_i \qquad R_i={K\over
 %N}\wvec_i\T \wvec_i^0
\reff={1\over K} \sum_{i=1}^K f(R_i) \qquad f(\cdot)={2\over\pi}\arcsin(\cdot)
 \]
 in terms of the overlaps of the student and teacher hidden unit
weight vectors, $R_i=(K/N)\wvec_i\T \wvec_i^0$.
In the thermodynamic limit, the $R_i$ are
self-averaging, \ie\ their values for a specific teacher, training set
and student from the Gibbs distribution are identical to their averages
with probability one. These averages can be obtained from a replica
calculation of the average entropy $S$ as a function of
 $\alpha$, following the calculations in
Refs.~\cite{Seungetal92b,Schwarzeetal92}.
We use the assumption of replica symmetry, which is believed to be exact
for the case of noise free training data~\cite{Schwarzeetal92}. The
replica calculation involves, in addition to the $R_i$, the overlap parameters
($\mu<p$)
 \[
q_i^p=(K/N)(\bar\wvec_i^p)^2 \qquad q_i^{\mu p}=(K/N)(\bar\wvec_i^p)\T
\bar\wvec_i^\mu
 %\label{qmup_definition}
 \]
 where $\bar\wvec_i^p=\av{\wvec_i}{P(\wvec|\thn)}$ and similarly for
$\bar\wvec_i^\mu$.
 %=\av{\wvec_i}{P(\wvec|\th{\mu})}$ ($\mu<p$).
 The $q_i^{\mu p}$ arise from the average over the ($\mu+1$)-th of the
$p$ training examples as the overlaps of the committee members which
determine the selection of this example with the students trained on all
$p$ examples.  The $q_i^p$ can be determined from saddle point
equations, whereas the $q_i^{\mu p}$ have to be determined
independently.  However, given the assumption of self-averaging of all
overlap parameters, it can be shown that $q_i^{\mu p}=q_i^\mu$ in the
case considered here~\cite{Sollich95b}.  This relation, which is proved
by induction from the case $p=\mu+1$, can be explained intuitively as
follows.  Given the first $\mu$ training examples $\th{\mu}$, the
teacher can be anywhere in the corresponding version space
$\cal{V}^\mu$.  Considering an average over all possible sets of
training examples $\mu+1\ldots p$ produced by teachers in $\cal{V}^\mu$,
the student is therefore equally likely to end up in any part of
$\cal{V}^\mu$ after having been trained on the whole training set
$\thn$.
 %the overlaps of two students trained on $\mu$ and
 %$p$ examples, respectively, are the same as the overlaps of two students
 %both trained on $\mu$ examples.

 %Following~\cite{Schwarzeetal92,Engeletal92}, we assume symmetry between the
 We assume symmetry between the
hidden units~\cite{Schwarzeetal92}, \ie\ $q_i^p=q$, $q_i^{\mu
p}=q_i^\mu=q(\alpha')$ ($\alpha'=\mu/N$) and $R_i=R$. The calculation, details
of which will be reported elsewhere~\cite{Sollich95b}, can be
further simplified by exploiting the
relation $R=q$, which expresses the symmetry between teacher and student
 %(see, \eg~\cite{Seungetal92b,Watkinetal93}). One then obtains the
 (see, \eg\ Ref.~\cite{Seungetal92b}). One then obtains the
normalized entropy $s=S/N$ (apart from an
additive constant, which we fix such that $s=0$ at $\alpha=0$)
as the saddle point of
 \$[
 %s= {\rm extr}_q\ \  \left\{
\half(q+\ln(1-q))+
2\int_0^\alpha \!d\alpha' \int\!Dz\ H(\gamma z) \ln H(\gamma z)
 %\right\}
\label{tcm_entropy}
 \$]
 with respect to $q$, where
 % \[
 %\gamma=\sqrt{{\qeff-\qeff(\alpha') \over 1-\qeff}}
 % %\gamma=[(\qeff-\qeff(\alpha'))/(1-\qeff)]^{1/2}
 % \]
$\gamma=[(\qeff-\qeff(\alpha'))/(1-\qeff)]^{1/2}$
 and $\qeff=f(q)$, $\qeff(\alpha')=f(q(\alpha'))$.
We have also used the shorthand
$Dz=dz \exp(-\half z^2)/\sqrt{2\pi}$ and
$H(z)=\int_z^\infty Dx$.
 %The effect of query learning enters
 %in~\(tcm_entropy) only through $\qeff(\alpha')$, and by replacing
 %$\qeff(\alpha')\to 0$, the known result for random
 %examples~\cite{Schwarzeetal92} is recovered.
 Differentiating~\(tcm_entropy) with respect to $\alpha$, one verifies that
$ds/d\alpha=-\ln 2$ as expected for minimum entropy queries
(the large committee limit $k\to\infty$ has already been
taken)~\cite{entropy_comes_out_right_footnote}.
\begin{figure}[thb]
\vspace*{-6mm}
\begin{center}
\begin{minipage}{8.5cm}
\begin{center}
\leavevmode
\epsfxsize = 8cm
\epsffile{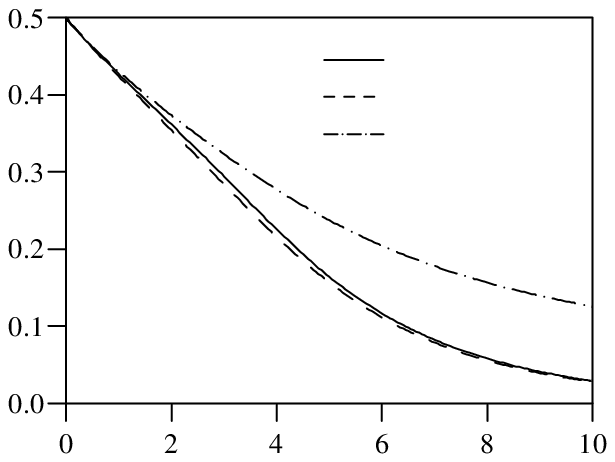}

\begin{picture}(50,0)

%\put(-9.5,47.3){\makebox(0,0){\normalsize $N_c(\alpha,\lambda)$}}
 %\put(28.5,54){\makebox(0,0){\normalsize (a)}}
 %\put(-6,39){\makebox(0,0){\normalsize $\epsg$}}
\put(-7,50.4){\makebox(0,0){\normalsize (a)}}
\put(-6,31.0){\makebox(0,0){\normalsize $\epsg$}}

\put(28.4,6.3){\makebox(0,0){\normalsize $\alpha$}}

\put(34,47){
\begin{picture}(0,0)(-0.8,0.63)
\put(0,0){\makebox(0,0)[l]{\normalsize exact}}
\put(0,-3.73){\makebox(0,0)[l]{\normalsize constructive}}
\put(0,-7.46){\makebox(0,0)[l]{\normalsize random}}
\end{picture}
}

\end{picture}
\end{center}
\end{minipage}
\vspace*{-0mm}
\begin{minipage}{8.5cm}
\begin{center}
\vspace*{-5mm}
\leavevmode
\epsfxsize = 8cm
\epsffile{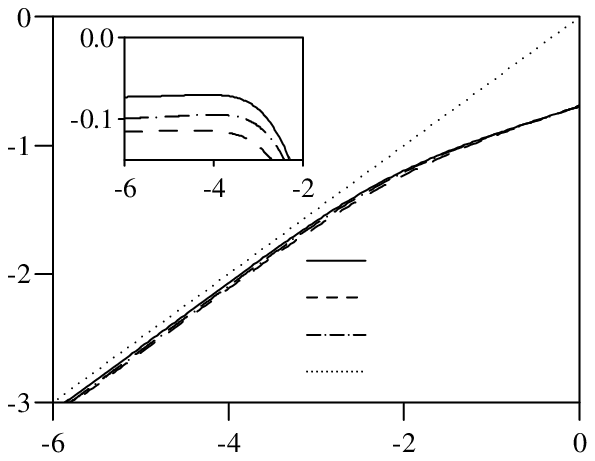}

\begin{picture}(50,0)

%\put(-9.5,47.3){\makebox(0,0){\normalsize $N_c(\alpha,\lambda)$}}
%\put(27,55){\makebox(0,0){\normalsize $N_c(\alpha,\lambda)$}}

 %\put(28.5,54){\makebox(0,0){\normalsize (b)}}
 %\put(-5,44.5){\makebox(0,0){\normalsize $\ln\epsg$}}
 \put(-6,50.55){\makebox(0,0){\normalsize (b)}}
\put(-5,31.5){\makebox(0,0){\normalsize $\ln\epsg$}}

%\put(28.4,6.3){\makebox(0,0){\normalsize $\alpha$}}

\put(29,6.3){\makebox(0,0){\normalsize $s$}}

\put(30,26.3){
\begin{picture}(0,0)(-4,0.2)
\put(0,0){\makebox(0,0)[l]{\normalsize exact}}
\put(0,-3.73){\makebox(0,0)[l]{\normalsize constructive}}
\put(0,-7.46){\makebox(0,0)[l]{\normalsize random}}
\put(0,-11.60){\makebox(0,0)[l]{\normalsize $\ln\epsg=\half s$}}
\end{picture}
}

\end{picture}
\end{center}
\end{minipage}
\end{center}
\vspace{-0.7cm}
\caption{(a) Generalization error $\epsg$ as a function of the normalized
number of examples, $\alpha$, for {\em exact} minimum entropy
queries, queries as selected by {\em constructive} algorithm,
and {\em random} examples. (b) Log generalization error $\ln\epsg$ vs.
entropy $s$, for the same three cases.
 %The thin full line corresponds to $\ln\epsg=\half s$.
For both queries and random examples,
$\ln\epsg\approx \half s$ for large negative values
of $s$ (corresponding to large $\alpha$). The very small separation
between the curves is more clearly seen in the inset, which shows
$\ln\epsg-\half s$ vs.~$s$.
 \label{fig:tcm_generr_exact_msse}
 \label{fig:tcm_generr_appr_msse}
 \label{fig:tcm_generr_entropy_rexs}
}

\end{figure}
Solving the saddle point equation
 %(which is an integral equation for
 %$q(\alpha)$)
 numerically, we obtain the average generalization error as
plotted in Figure~\ref{fig:tcm_generr_exact_msse}(a). For large $\alpha$,
we find that $\epsg\propto\exp(-c\alpha)$ with $c=\half\ln 2$, which can also
be confirmed analytically from~\(tcm_entropy).
This exponential decay of the generalization error $\epsg$ with $\alpha$
provides a marked improvement over the $\epsg\propto 1/\alpha$ decay
achieved by random examples~\cite{Schwarzeetal92}.  The effect of
minimum entropy queries is thus similar to what is observed for
a binary perceptron learning from a binary perceptron teacher,
but the decay constant $c$ is only half of that for the binary
perceptron~\cite{Seungetal92b}.  This means that asymptotically, twice
as many examples are needed for a TCM as for a binary perceptron
(when learning from a teacher with the respective architecture) to
achieve the same generalization performance, in agreement with the
corresponding result for random examples.
 %~\cite{Schwarzeetal92}.
 Since
in both networks, due to the binary nature of their outputs, minimum
entropy queries lead to an entropy $s=-\alpha\ln 2$, we can
also conclude that the large $\alpha$ relation $s\approx\ln\epsg$ for the
binary
perceptron~\cite{Seungetal92b} has to be replaced by
$s\approx\ln\epsg^2$ for the tree committee machine.  This relation
should hold independently of whether one is learning from
queries or from random examples.  We have confirmed this by calculating
the entropy for learning from random examples and comparing with the
corresponding generalization error,
 %~\cite{Schwarzeetal92},
 as shown in Figure~\ref{fig:tcm_generr_entropy_rexs}(b).
 %
 %\begin{figure}[t]
 %%
 %\begin{center}
 %%
 %\leavevmode
 %%\epsfxsize = 6cm
 %\epsffile{/home/pkso/project/manna/tcm_log_generrs_vs_entropy.ps}
 %
 %\begin{picture}(50,0)
 %
 %%\put(-9.5,47.3){\makebox(0,0){\normalsize $N_c(\alpha,\lambda)$}}
 %%\put(27,55){\makebox(0,0){\normalsize $N_c(\alpha,\lambda)$}}
 %\put(-5,46.5){\makebox(0,0){\normalsize $\ln\epsg$}}
 %
 %%\put(28.4,6.3){\makebox(0,0){\normalsize $\alpha$}}
 %
 %\put(32.6,6.3){\makebox(0,0){\normalsize $s$}}
 %
 %\put(30,26.3){
 %\begin{picture}(0,0)
 %\put(0,0){\makebox(0,0)[l]{\normalsize exact}}
 %\put(0,-3.73){\makebox(0,0)[l]{\normalsize constructive}}
 %\put(0,-7.46){\makebox(0,0)[l]{\normalsize random}}
 %\end{picture}
 %}
 %
 %\end{picture}
 %%
 %\end{center}
 %%
 %\vspace{-0.8cm}
 %%
 % \caption{Log generalization error $\ln\epsg$ vs. (normalized) entropy
 %$s$, for queries (exact/constructive algorithm) and random examples.
 %The thin full line corresponds to $\ln\epsg=\half s$.
 %For all three cases, $\ln\epsg\approx \half s$ for large negative values
 %of $s$ (corresponding to large $\alpha$).
 % \label{fig:tcm_generr_entropy_rexs}
 %}
 %
 %\end{figure}

 The above results are derived in the limit of a large number of hidden
units, $K\to\infty$.  For large but finite $K$ they can be shown to be
valid as long as the $O(1/K)$ correction to the generalization
error~\(tcm_generr), $(-1/2\pi K)\reff(1-\reff^2)^{1/2}$, remains
negligible, which holds for $\epsg\gg O(K^{-1/2})$.  In the opposite
regime $\epsg\ll O(K^{-1/2})$, \ie\ for higher $\alpha$, the
generalization error $\epsg\approx (K/8)^{1/2}(1-\reff)\propto
\arccos(R)$ has the same functional dependence on $R$ as for the binary
perceptron, due to the fact that its dominant contribution arises from
errors for which student and teacher only differ in the output of a
single hidden unit.  There is therefore a cross-over in the large
$\alpha$ dependence of $\epsg$ from TCM ($K\to\infty$) to binary
perceptron type behaviour around $\epsg=O(K^{-1/2})$.

 %\section{
 % %Approximate minimum entropy que\-ries}
 %Constructive query selection algorithm}
\label{sec:calcul_appr_msse_queries}

We now consider the practical realization of minimum entropy
queries in the TCM.  The query by committee approach, which in the limit
$k\to\infty$ is an exact algorithm for selecting minimum entropy
queries, filters queries from a stream of random inputs.  This leads to
an exponential increase of the query filtering time with the number of
training examples that have already been learned~\cite{Freundetal93}.
As a computationally cheap alternative we propose a simple algorithm for {\em
constructing} queries, which is based on the assumption of an approximate
decoupling of the entropies of the different hidden units, as follows.
Each individual hidden unit of a TCM can be viewed as a binary
perceptron.  The distribution $P(\wvec_i|\thn)$ of its weight vector $\wvec_i$
given a set of training examples $\thn$ has an entropy $S_i$ associated
with it, in analogy to the entropy~\(tcm_entropy_general) of the full
weight distribution $P(\wvec|\thn)$.  Our `constructive
algorithm' for selecting queries
then consists in choosing, for each new query $\xvec^{\mu+1}$,
the inputs $\xvec_i^{\mu+1}$ to the individual hidden units in such a way as
to maximize the decrease in their entropies $S_i$. This can be achieved
simply by choosing
each $\xvec_i^{\mu+1}$ to be orthogonal to $\bar\wvec_i^\mu$ (and
otherwise random, \ie\ according to $P_0(\xvec)$)~\cite{Sollich95b},
thus avoiding the time-consuming
filtering from a random input stream.  In practice, one would of course
approximate $\bar\wvec_i^\mu$ by an average of $2k$ (say) samples
from the Gibbs distribution $P(\wvec|\th{\mu})$; these samples would
have been needed anyway in the query by committee approach.

An analysis of the generalization performance achieved by this constructive
algorithm proceeds along the same line as the
calculation for exact minimum entropy queries.
Again restricting attention to the
limit $k\to\infty$, we find that the saddle point expression~\(tcm_entropy)
for the
normalized entropy $s$ still holds, but with $\gamma$ now given by
$\gamma=[a/(1-a)]^{1/2}$, $a=f([q-q(\alpha')]/[1-q(\alpha')])$.
 % \[
 %\gamma=\sqrt{{a\over 1-a}} \qquad a={2\over\pi}
 %\arcsin\left({q-q(\alpha')\over 1-q(\alpha')}\right)
 % %\gamma=[a/(1-a)]^{1/2} \quad a=(2/\pi)
%%\arcsin[(q-q(\alpha'))/(1-q(\alpha'))]
 % \]
 Differentiating~\(tcm_entropy) with this replacement with respect to
$\alpha$, we find again that $ds/d\alpha=-\ln 2$, which means that in
the thermodynamic limit that we consider, queries selected to minimize
the individual hidden units' entropies also minimize the overall entropy
of the TCM.  This may seem surprising at first; heuristically, however,
one can argue that for a large number of hidden units $K$, the
correlations in the Gibbs distribution between the hidden unit weight
vectors must be weak, and may indeed become negligible in the
$K\to\infty$ limit considered here.  The generalization performance
achieved by the constructive query algorithm, shown in
Figure~\ref{fig:tcm_generr_appr_msse}(a), is actually slightly superior to
that of exact minimum entropy queries as calculated in the
previous section.  This decrease in generalization error, although
slight (about 4\% for large $\alpha$), exemplifies the fact that while
decrease in entropy and in generalization error are normally
correlated, there is no exact one-to-one relationship between them
(compare the discussion in Ref.~\cite{Sollich94}).  Query selection
algorithms which achieve the same entropy decrease can therefore lead to
different generalization performance.

\label{sec:tcm_conclusions}

 %We have used the tools of statistical mechanics to analyse query
 %learning for minimum entropy in large tree-committee machines
 %(TCM).  For the noise free, perfectly learnable scenario that we have
 %considered, the generalization error $\epsg$ decays exponentially with
 %the normalized number of training examples $\alpha$, which is a
 %significant improvement over learning from random examples, for which
 %$\epsg\propto 1/\alpha$ for large $\alpha$.  Comparing with results for
 %query learning in the binary perceptron, the decay constant $c$ in
 %$\epsg\propto\exp(-c\alpha)$ turns out to be half as large in the TCM,
 %and this implies that the relationship between entropy $s$ and
 %generalization error is $s\approx\ln\epsg^2$ in the TCM, rather than
 %$s\approx\ln\epsg$ as in the binary perceptron.

 %Comparing our results with those for the binary perceptron,
 We have
found above a modification of the relationship between entropy $s$ and
generalization error $\epsg$ from $s\approx\ln\epsg$ for the binary
perceptron to $s\approx\ln\epsg^2$ for the TCM, and a corresponding
change of the decay constant $c$ in the asymptotic behaviour of the
generalization error $\epsg\propto\exp(-c\alpha)$.  This leads to the
interesting question of the value of $c$ in more general multi-layer
neural networks, and in particular its dependence on the number of
hidden units $K$.  The bound in Ref.~\cite{Freundetal93}, derived for the
$k=1$ query by committee algorithm, implies a lower bound on $c$ which
scales inversely with the VC-dimension~\cite{Vapniketal71} of
 % VC-dimension of
the class of networks considered.  Taking the storage capacity of a
network as a coarse measure of its VC-dimension, one would then conclude
from existing bounds~\cite{Mitchisonetal89}
that $c$ could be as small as $O(1/\ln K)$ for large $K$.
However, the existing results for the capacity of particular networks
like the TCM
 %(which was conjectured to be finite for $K\to\infty$
 %in~\cite{Opper94} on the basis of the results of~\cite{Schwarzeetal92},
 %but estimated to be infinite in the same limit in~\cite{Barkaietal92})
 %
 are not unambiguous enough to decide whether
 %
 %the VC-dimension of
 %
 realistic
networks would saturate this bound.
 %bound~\cite{Mitchisonetal89}.
 Furthermore, it has been argued
previously~\cite{Opper94} that both the input space dimension {\em and}
 % previously that both the input space dimension {\em and}
 the VC-dimension determine the $\alpha$-dependence of the generalization
error. Replacing the VC-dimension in the
bound in Ref.~\cite{Freundetal93} with the input space dimension,
one would then obtain a $c$ of $O(1)$
independently of $K$. More theoretical work is clearly needed to clarify
these questions.
 %
 %It may therefore be possible to replace
 % %if this is
 % %significantly smaller,
 % and this would yield a $c$ of $O(1)$
 %independently of $K$. Further theoretical work is clearly needed to
 %clarify the above questions.

With regard to the practical application of query learning in realistic
multi-layer neural networks, the
results we have obtained for a constructive query algorithm based on
 %In the second part of the paper we have analysed a computationally cheap
 %algorithm for constructing (rather than filtering) approximate maximum
 %information gain queries, based on
 the assumption of a decoupling of the
entropies of individual hidden units are encouraging.
 %We have found that this
 %constructive algorithm actually achieves slightly better generalization
 %performance than exact minimum entropy queries.  This result is
 %particularly encouraging considering the practical application of query
 %learning in more complex multi-layer networks.
For example, the proposed
constructive algorithm
 %discussed in
 %section~\ref{sec:calcul_appr_msse_queries}
 can be modified for query learning in a fully-connected committee
machine (where each hidden unit is connected to all the inputs), by
simply choosing each new query to be orthogonal to the subspace spanned
by the average weight vectors of {\em all} $K$ hidden units.  As long as
$K$ is much smaller than the input dimension $N$, and assuming that for
large enough $K$ the approximate decoupling of the hidden unit entropies
still holds for fully connected networks, one would expect this
algorithm to yield a good approximation to minimum entropy
queries~\cite{permutation_breaking_footnote}.
It is an open question whether this conclusion would also hold for a
general two-layer network with threshold units (where
 % in contrast to the committee machine,
the hidden-to-output weights are also free
parameters), which can approximate a large class of input-output
mappings. We are currently investigating these issues in order to assess
whether the significant
improvements in generalization performance achieved by minimum entropy
queries can be made available, in a computationally cheap manner, for
learning in realistic binary output multi-layer neural networks.


\begin{thebibliography}{10}
\bibitem{entropy_comes_out_right_footnote}
The fact that the entropy is predicted correctly lends further support to the
  claim that the assumption of replica symmetry is correct for the problem
  considered here.
\bibitem{permutation_breaking_footnote}
 To make the algorithm work once the permutation symmetry between
hidden units is broken, one would of course have to restrict all weight
space averages to one of the $K!$ `ergodic' weight space sectors.
\end{thebibliography}
\end{document}